\documentclass{eptcs}

\usepackage{amssymb}
\usepackage{proof}
\usepackage{stmaryrd}
\usepackage{url}
\usepackage{code}

\title{Resumption-based big-step and small-step interpreters\\ for While with interactive I/O}

\def\titlerunning{Resumption-based big-step and small-step total interpreters for While with interactive I/O}

\author{
Keiko Nakata 
\institute{Institute of Cybernetics at Tallinn University of Technology, %\\
Akadeemia tee 21, EE-12618 Tallinn, Estonia} 
\email{keiko@cs.ioc.ee}
}

\def\authorrunning{K.~Nakata}

\begin{document}

\maketitle

\begin{abstract}
  In this tutorial, we program big-step and small-step total
  interpreters for the While language extended with input and output
  primitives. While is a simple imperative language consisting of
  skip, assignment, sequence, conditional and loop. We first develop
  trace-based interpreters for While. Traces are potentially infinite
  nonempty sequences of states. The interpreters assign traces to
  While programs: for us, traces are denotations of While programs.
  The trace is finite if the program is terminating and infinite if
  the program is non-terminating. However, we cannot decide (i.e.,
  write a program to determine), for any given program, whether its
  trace is finite or infinite, which amounts to deciding the halting
  problem.  We then extend While with interactive input/output
  primitives. Accordingly, we extend the interpreters by generalizing
  traces to resumptions. 
  
  The tutorial is based on our previous work with T. Uustalu on reasoning
  about interactive programs in the setting of constructive type theory.
\end{abstract}

\section{Introduction}

\emph{Interactive} programs are those programs that take inputs, do some
computation, output results, and iterate this cycle possibly infinitely. 
Operating systems and data base systems are typical examples. 
They are important programs and have attracted formal study
to guarantee their correctness/safety. For instance, 
a web application should protect confidentiality of the data it  processes 
in interaction with possibly untrusted agents, 
and a certified compiler should preserve 
input/output behavior of the source program in the compiled code.
These works call for formal semantics of interactive programs. 

In our previous work, we presented a constructive account of
interactive input-output resumptions\footnote{The word `resumption' is
  sometimes reserved for denotations of parallel threads. We apply it
  more liberally to datastructures recording evolution in small
  steps. This usage dates back to Plotkin \cite{Plo:dom} and was
  reinforced by Cenciarelli and Moggi~\cite{CM:synmds}.}, their
important properties, such as weak bisimilarity and responsiveness
(a program always eventually performs input or output unless it
terminates), and several coinductive operational semantics for interactive
programs. Our operational semantics are resumption-based.  A
resumption is roughly a tree representing possible runs of a
program. The tree branches on inputs, each edge corresponding to each
possible input, and has infinitely deep paths if the program may
diverge. The development has been fully formalized in the Coq proof
assistant.

This tutorial serves to deliver the central ingredients behind our
coinductive semantics by programming resumption-based interpreters for
interactive programs in Haskell.  No knowledge of Coq or coinduction
is assumed. We use Haskell as we can naturally build and manipulate
infinite objects. However, one can also work with OCaml using the lazy
and force primitives.

The tutorial starts by programming trace-based interpreters for 
the While language.
Traces, defined coinductively, are possibly infinite nonempty 
sequences of states. 
The interpreters assign traces to programs 
(to be more precise, to statement-state pairs),
recording the states that program runs go through. 
The trace is finite if the program run is terminating 
and it is infinite if the program run is non-terminating. 
Unlike the standard state-based interpreter for While, 
our trace-based interpreters
are total: they return traces for \emph{all} programs. 
We will develop both big-step and small-step interpreters, 
which are provably equivalent constructively. (The proof of the 
equivalence is beyond the scope of the tutorial.)

We then extend While with interactive input/output primitives
and generalize traces to resumptions. Accordingly, 
we extend our big-step and small-step interpreters, which now 
assign resumptions to interactive While programs. 
Once one learns trace-based interpreters for While,
this step is rather straightforward. 

This tutorial is based on our previous work~\cite{nonterm,interact}.
The whole development in the tutorial 
is portable to Coq with minor syntax adjustment.
We will be explicit in whether a datatype is defined inductively 
(\verb#Inductive# in Coq's vernacular) or
coinductively (\verb#CoInductive#) 
and whether a function is defined by recursion 
(\verb#Fixpoint#) or by
corecursion (\verb#CoFixpoint#), 
although these distinctions are not visible when one 
works in Haskell. 

The accompanying Haskell code can be downloaded 
from\\ \centerline{\url{http://cs.ioc.ee/~keiko/code/dsl11.tgz}.}

\section{Trace-based interpreters for While}

\emph{States} are functions from variables to values\footnote{The Haskell code,
in particular the notations, 
are adapted from the lecture material for a programming language semantic
course by T. Uustalu.}.
We represent both \emph{variables} and \emph{values} by integers:
\begin{code}
type Var = Integer
type Val = Integer
type State = Var -> Val
\end{code}
Looking up a variable in a state is then simply function application: 
\begin{code}
lkp :: Var -> State -> Val
lkp x s = s x
\end{code}
We also define update \verb#upd x v s# of a state \verb#s# 
at a variable \verb#x# by a value \verb#v#:
\begin{code}
upd :: Var -> Val -> State -> State
upd x v s = \\y -> if x == y then v else s y
\end{code}

\smallskip

The syntax of arithmetic expressions is defined inductively by
\begin{code}
data AExp = N Integer | V Var 
          | AExp :+ AExp | AExp :- AExp | AExp :* AExp
\end{code}
The function \verb#aexp a s# evaluates an expression \verb#a# 
in a state \verb#s#.
It is defined by recursion over the syntax of expressions: 
\begin{code}
aexp :: AExp -> State -> Integer
aexp (N z) \_      = z
aexp (V x) s      = lkp x s
aexp (a0 :+ a1) s = aexp a0 s + aexp a1 s
aexp (a0 :- a1) s = aexp a0 s - aexp a1 s
aexp (a0 :* a1) s = aexp a0 s * aexp a1 s
\end{code}
Similarly, we define boolean expressions and an evaluator for them:
\begin{code}
data BExp = TT | FF | AExp :== AExp | AExp :<= AExp 
          | Not BExp | BExp :&& BExp | BExp :|| BExp
\-
bexp :: BExp -> State -> Bool
bexp TT \_ = True
bexp FF \_ = False
bexp (a0 :== a1) s = aexp a0 s == aexp a1 s
bexp (a0 :<= a1) s = aexp a0 s <= aexp a1 s
bexp (Not b) s = not (bexp b s)
bexp (a0 :&& a1) s = bexp a0 s && bexp a1 s
bexp (a0 :|| a1) s = bexp a0 s || bexp a1 s
\end{code}

\smallskip

The syntax of the While language is defined inductively by
\begin{code}
data Stmt = Skip | Stmt :\\ Stmt | Var := AExp
          | If BExp Stmt Stmt | While BExp Stmt
\end{code}

Our interpreters for While assign \emph{trace}s to While programs.
Traces are possibly infinite nonempty sequences of states.
They are defined coinductively by
\begin{code}
data Trace = Nil State | Delay State Trace
\end{code}
A trace may be finite or infinite. But we cannot decide, 
for any given trace \textit{t}, whether \textit{t} is finite or not.
In other words, we cannot write a (total) function that returns true when 
the trace is finite and returns false otherwise. (Why?)

\subsection{Big-step interpreter}

The standard state-based interpreter for While is partial: given
a statement and an initial state, it returns the final state 
if running the statement from the initial state is terminating.
The interpreter diverges if the statement runs forever. 
When one works in a setting where only 
total functions are definable, e.g., within the logic of Coq, the state-based
interpreter cannot be defined constructively, 
as this would require deciding the halting problem, an instance of
the Principle of the Excluded Middle.
Working with traces has the benefit that we do not need to decide: 
any statement and initial state uniquely determine some trace and we do not
have to know whether this trace is finite for infinite. 

Our trace-based big-step interpreter \verb#eval# 
takes a statement and an initial state
and returns a trace. It is defined by recursion over the syntax
of statements:
\begin{code}
eval :: Stmt -> State -> Trace
eval Skip s = Nil s
eval (stmt0 :\\ stmt1) s = seque (eval stmt1) (eval stmt0 s)
eval (x := a) s = Delay s (Nil (upd x v s)) where v = aexp a s
eval (If b stmt0 stmt1) s =
    if bexp b s then 
      Delay s (eval stmt0 s) 
    else Delay s (eval stmt1 s)
eval (While b stmt0) s =
    if bexp b s then
      Delay s (loop (eval stmt0) (bexp b) s)
    else Delay s (Nil s)
\end{code}

We consider \verb#Skip# to be terminal, or it does not take
time to run \verb#Skip#, so the interpreter 
returns a singleton consisting of the initial state.
The trace for assignment \verb#x := a# is a doubleton: it consists 
of the initial state and the final state obtained by
updating the initial state \verb#s# at \verb#x# by the value of \verb#a# 
in \verb#s#.
For sequence \verb#stmt0 :\ stmt1#, 
we use an auxiliary function \verb#seque#, defined by corecursion by
\begin{code}
seque :: (State -> Trace) -> Trace -> Trace
seque k (Nil s) = k s
seque k (Delay s t) = Delay s (seque k t)
\end{code}
The idea is that we first run the statement \verb#stmt0# from the initial state,
then \verb#stmt1# is run from the last state of the trace
produced by the run of \verb#stmt0# (if the last state exists).
In particular, \verb#stmt1# will not be run at all if
running \verb#stmt0# from the initial state is nonterminating:
then the trace for \verb#stmt0# is infinite and we never get to its
last state, from where \verb#stmt1# will be run.

For conditional, the appropriate branch is run depending
on whether the boolean guard evaluates to true or false.
The trace contains one additional delay to the trace corresponding to
the run of the branch, accounting for the time taken to evaluate 
the guard. 

For while, we use an auxiliary function \verb#loop#
defined by mutual corecursion together with \verb#loopseq#:
\begin{code}
loop :: (State -> Trace) -> (State -> Bool) -> State -> Trace
loop k p s = 
    if p s then
        case k s of
          Nil s' -> Delay s' (loop k p s')  
          Delay s' t -> Delay s' (loopseq k p t)
    else Nil s
loopseq :: (State -> Trace) -> (State -> Bool) -> Trace -> Trace
loopseq k p (Nil s) = Delay s (loop k p s)
loopseq k p (Delay s t) = Delay s (loopseq k p t)
\end{code}
The function \verb#loop#
takes three arguments: \verb#k# for evaluating the loop body from a state;
\verb#p# for testing the boolean guard on a state; and a state 
\verb#s#, which
is the initial state. \verb#loopseq# takes a trace, the initial
trace, instead of a state, as the third argument.  The two functions
work as follows. \verb#loop# takes care of repeating of the loop body,
once the guard of a while loop has been evaluated.  It analyzes the
result and, if the guard is false, then the run of the loop
terminates.  If it is true, then the loop body is evaluated by
calling \verb#k#. \verb#loop# then constructs the trace of the loop body by
examining the result of \verb#k#.  If the loop body does not augment the
trace, which can only happen, if the loop body is a sequence of
\verb#Skip#s, a new round of repeating the loop body is started by a
recursive call to \verb#loop#.  
If the loop body augments the trace, the new round
is reached by reconstruction of the trace of the current repetition
with \verb#loopseq#. On the exhaustion of this trace, \verb#loopseq#
recursively calls \verb#loop#.

As a Haskell program, one might not find 
the definitions of \verb#loop# and \verb#loopseq# most intuitive.
Indeed they are arranged so that (co)recursive calls to 
\verb#loop# and \verb#loopseq# are ``guarded'' by a \verb#Delay# constructor.
This way, Coq guarantees these functions are productive, 
as required by the logic of Coq. 

\medskip

Some design decisions we have made are that \verb#Skip# does not grow a
trace, so we have
\begin{code}
eval Skip s = Nil s
\end{code}
But an assignment and
testing the guard of an if- or while-statement contribute a state,
i.e., constitute a small step, e.g., we have 
\begin{code}
eval (x := 17) s = Delay s (Nil (upd x 17 s))
eval (While FF Skip) s = Delay s (Nil s)
\end{code}
and 
\begin{code}
eval (While TT Skip) s = Delay s (Delay s (Delay s (...)))
\end{code}
This is good for several reasons. First, we have that \verb#Skip# is the
identity of sequential composition, i.e., the semantics does not
distinguish \textit{stmt}, \verb#Skip :\# \textit{stmt} and 
\textit{stmt} \verb#:\ Skip# for any statement \textit{stmt}.  Second, we
get a notion of small steps that fully agrees with the textbook-style
small-step interpreter given in the next section. The third and most
important outcome is that any while-loop always progresses, because
testing of the guard is a small step.  Another option would be to
regard testing of the guard to be instantaneous, but take leaving the
loop body, or a backward jump in terms of low-level compiled code, to
constitute a small step. But then we would not agree to the textbook
small-step interpreter. 

\subsection{Small-step interpreter}

We proceed to an equivalent small-step interpreter for While.
It is based on an option-returning one-step reduction function 
\verb#red#, defined by recursion over the syntax of statements:
\begin{code}
red :: Stmt -> State -> Maybe (Stmt, State)
red Skip s = Nothing
red (x := a) s = Just (Skip, upd x v s) where v = aexp a s
red (stmt0 :\\ stmt1) s = 
    case red stmt0 s of
      Just (stmt0', s') -> Just (stmt0' :\\ stmt1, s')
      Nothing -> red stmt1 s
red (If b stmt0 stmt1) s = 
    if bexp b s then
      Just (stmt0, s)
    else Just (stmt1, s)
red (While b stmt0) s =
    if bexp b s then
      Just (stmt0 :\\ While b stmt0, s)
    else Just (Skip, s)
\end{code}
The function \verb#red# returns \verb#Nothing#
if the given statement is terminal, otherwise
it one-step reduces the given statement from the given state 
and returns the resulting statement-state pair. 
Then the small-step interpreter \verb#norm# is defined by corecursion
by repeatedly calling \verb#red#:
\begin{code}
norm :: Stmt -> State -> Trace
norm stmt s =
    case red stmt s of
      Nothing -> Nil s
      Just (stmt', s') -> Delay s (norm stmt' s')
\end{code}

One can in fact prove that the big-step and small-step 
interpreters are equivalent:
for any statement \textit{stmt} and state \textit{s},
\verb#eval# \textit{stmt} \textit{s} and \verb#norm# \textit{stmt} 
\textit{s}
returns equal traces. The proof is found in~\cite{nonterm}, 
which is however beyond the scope of this tutorial. 

\section{Resumption-based interpreters for While with interactive I/O}

We now extend While with interactive input/output primitives.
The new syntax for statements is defined inductively by 
\begin{code}
data Stmt = Skip | Stmt :\\ Stmt | Var := AExp
          | If BExp Stmt Stmt | While BExp Stmt
          | Input Var | Output AExp
\end{code}
The statement \verb#Input x# reads an input value and stores it
at the variable \verb#x#.
The statement \verb#Output a# evaluates the expression \verb#a#
in the current state and outputs the resulting value. 

To account for interactive input/output, we generalize traces to
\emph{resumptions}. Informally, a resumption is a datastructure that
captures all possible evolutions of a configuration (a statement-state
pair), a computation tree branching according to the external
non-determinism resulting from interactive input.\footnote{There are
  alternatives. We could have chosen to work, e.g., with functions
  from streams of input values into traces, i.e., computation paths.}

Resumptions are defined coinductively by 
\begin{code}
data Res = Ret State | In (Val -> Res) | Out (Val, Res) | Delay Res 
\end{code}
so a resumption either has terminated with some final state,
\verb#Ret s#, takes an input value \verb#v# and evolves into a new
resumption \verb#f v#, \verb#In f#, outputs a value \verb#v# and 
evolves into \verb#r#, \verb#Out (v, r)#, or performs an internal 
action (observable at best as a delay) and becomes \verb#r#,
\verb#Delay r#.

Here are some examples of resumptions, defined by corecursion:
\begin{code}
bot :: Res
bot = Delay bot
rep :: Val -> Res
rep v = Delay (Delay (Out (v, rep v)))
rep' :: Val -> Res
rep' v = Delay (Out (v, rep' v))
echo :: State -> Res
echo s = In (\\v -> Delay (if v == 0 then Out (v, echo s) else Ret s))
echo' :: Res
echo' = In (\\v -> Delay (if v == 0 then Out (v, echo') else bot))
\end{code}
\verb#bot# represents a resumption that silently diverges.
\verb#rep v# outputs a value \verb#v# forever. \verb#rep' v# is similar
but has shorter latency. Both \verb#echo# and \verb#echo'# echo input
interactively; the former terminates when the input is 0,
whereas the latter diverges in this situation. 

\subsection{Big-step interpreter}

Extending the big-step interpreter for While to handle input/output primitives
is straightforward. The new interpreter is given in figure~\ref{fig:bigstepres}.
Input and output statements evaluate to corresponding resumptions
that perform input or output actions and terminate thereafter. 
The functions \verb#seque#, \verb#loop# and \verb#loopseq# are
extended in an expected way. 

\begin{figure}[t]
\begin{code}
eval :: Stmt -> State -> Res
eval Skip s = Ret s
eval (stmt0 :\\ stmt1) s = seque (eval stmt1) (eval stmt0 s)
eval (x := a) s = Delay (Ret (upd x v s)) where v = aexp a s
eval (If b stmt0 stmt1) s =
    if bexp b s then 
      Delay (eval stmt0 s) 
    else Delay (eval stmt1 s)
eval (While b stmt0) s =
    if bexp b s then
      Delay (loop (eval stmt0) (bexp b) s)
    else Delay (Ret s)
eval (Input x) s = In (\\v -> Ret (upd x v s))
eval (Output a) s = Out (v, Ret s)  where v = aexp a s 
\-
seque :: (State -> Res) -> Res -> Res
seque k (Ret s) = k s
seque k (In f) = In (\\v -> seque k (f v))
seque k (Out (v, r)) = Out (v, seque k r)
seque k (Delay r) = Delay (seque k r)
\-
loop :: (State -> Res) -> (State -> Bool) -> State -> Res
loop k p s = 
    if p s then
        case k s of
          Ret s' -> Delay (loop k p s')  
          In f -> In (\\v -> loopseq k p (f v))
          Out(v, r) -> Out (v, r') where r' = loopseq k p r
          Delay r -> Delay r' where r' = loopseq k p r
    else Ret s

loopseq :: (State -> Res) -> (State -> Bool) -> Res -> Res
loopseq k p (Ret s) = Delay (loop k p s)
loopseq k p (In f) = In (\\v -> loopseq k p (f v))
loopseq k p (Out(v, r)) = Out (v, loopseq k p r)
loopseq k p (Delay r) = Delay (loopseq k p r)
\end{code}
\caption{Resumption-based big-step interpreter for While with I/O}
\label{fig:bigstepres}
\end{figure}

\subsection{Small-step interpreter}

To define an equivalent small-step interpreter for the interactive While, 
we introduce labeled configurations, defined inductively by: 
\begin{code}
data Lconf = Ret\_ State | In\_ (Stmt, Val -> State) 
           | Out\_ (Val, Stmt, State) | Delay\_ (Stmt, State)
\end{code}

The one-step reduction function \verb#red# for the interactive While
returns a labeled configuration, given a statement-state pair.
It is defined by recursion over the syntax of statements by
\begin{code}
red :: Stmt -> State -> Lconf
red Skip s = Ret\_ s
red (x := a) s = Delay\_ (Skip, upd x v s) where v = aexp a s
red (stmt0 :\\ stmt1) s = 
    case red stmt0 s of
      Ret\_ s' -> red stmt1 s'
      In\_ (stmt0', f) -> In\_ (stmt0' :\\ stmt1, f)
      Out\_ (v, stmt0', s') -> Out\_ (v, stmt0' :\\ stmt1, s')
      Delay\_ (stmt0', s') -> Delay\_ (stmt0' :\\ stmt1, s')
red (If b stmt0 stmt1) s =
    if bexp b s then
      Delay\_ (stmt0, s)
    else Delay\_ (stmt1, s)
red (While b stmt0) s =
    if bexp b s then
      Delay\_ (stmt0 :\\ While b stmt0, s)
    else Delay\_ (Skip, s)
red (Input x) s = In\_ (Skip, \\v -> upd x v s)
red (Output a) s = Out\_ (v, Skip, s)  where v = aexp a s
\end{code}
Then the small-step interpreter is again obtained by repeatedly
calling \verb#red#. It is defined by corecursion by
\begin{code}
norm :: Stmt -> State -> Res
norm stmt s = 
    case red stmt s of
      Ret\_ s' -> Ret s'
      In\_ (stmt', f) -> In (\\v -> norm stmt' (f v))
      Out\_ (v, stmt', s') -> Out (v, norm stmt' s')                        
      Delay\_ (stmt', s') -> Delay (norm stmt' s')
\end{code}

\subsection{Reasoning with resumptions}\label{sec:reasoning}

Resumptions are a syntax-free representation of the behavior of programs.
We can reason about the behavior of programs in terms of resumptions they
produce. In this subsection, we formalize two important properties
of resumptions, namely responsiveness and delay bisimilarity\footnote{We assume extensional equality on resumptions.}.
Informally, a resumption is responsive 
if it always eventually performs an input or output action unless it
terminates. Two resumptions are delay-bisimilar if they agree modulo
finite delays. 

A predicate $P$ on resumptions is \emph{responsive} if,
whenever $P\, r$ holds, then one of the following conditions holds:
\begin{enumerate}
\item $r$ = \verb#Delay#$^n$ (\verb#Res# $s$) for some $n$ and $s$;
\item $r$ = \verb#Delay#$^n$ (\verb#In# $f$) for some $n$ and $f$, and
for any value $v$, $P\, (f\, v)$;
\item $r$ = \verb#Delay#$^n$ (\verb#Out# ($v$, $r'$)) 
for some $n$ and $r'$ and $P\, r'$.
\end{enumerate}
where the notation \verb#Delay#$^n$ $r$
denotes $\overbrace{\mbox{\sf Delay} (... (\mbox{\sf Delay}}^n\, r)...)$.

A resumption $r$ is responsive if there is a responsive predicate
$P$ such that $P\, r$ holds. 

For instance, \verb#echo# \textit{s} given earlier in this section
is responsive for any \textit{s}, but \verb#echo'# is not.
(Take $P$ such that, for any $r$, $P\, r$ holds if
$r$ is either \verb#echo# \textit{s}, 
\verb#Delay# (\verb#Out# (\textit{v}, \verb#echo# \textit{s}))
or \verb#Delay# (\verb#Ret# \textit{s}).)

\medskip

A binary relation $R$, written in infix notation,
on resumptions is a \emph{(termination-sensitive) delay-bisimulation}~\cite{WeijlandPhD}
if, whenever $r_0\, R\, r_1$ holds, then one of the following conditions holds:
\begin{enumerate}
\item $r_0$ = \verb#Delay#$^n$ (\verb#Res# $s$)
and $r_1$ = \verb#Delay#$^{n'}$ (\verb#Res# $s$)
for some $n$ and $n'$;
\item $r_0$ = \verb#Delay#$^n$ (\verb#In# $f_0$) and
$r_1$ = \verb#Delay#$^{n'}$ (\verb#In# $f_1$)
for some $n, n', f_0$ and  $f_1$, 
and, for any value $v$, $(f_0\, v)~R~(f_1\, v)$.
\item $r_0$ = \verb#Delay#$^n$ (\verb#Out# ($v$, $r'_0$)) and 
$r_1$ = \verb#Delay#$^{n'}$ (\verb#Out# ($v$, $r'_1$))
for some $n, n', r'_0$ and $r'_1$, 
and $r'_0~R~r'_1$.
\item $r_0$ = \verb#Delay# $r'_0$ and
$r_1$ = \verb#Delay# $r'_1$ for some $r'_0$ and $r'_1$, and $r'_0~R~r'_1$.
\end{enumerate}

Two resumptions $r_0$ and $r_1$ are delay-bisimilar if 
there is a delay-bisimulation $R$ such that $r_0\, R\, r_1$.

For instance, \verb#rep# \textit{v} and \verb#rep'# \textit{v} 
are delay-bisimilar for any
\textit{v}. (Take $R$ such that, for any $r$ and $r'$,  $r\, R\, r'$ 
holds if $r$ = \verb#rep# \textit{v} and $r'$ = \verb#rep'# \textit{v}.)

\section{Some further reading}\label{sec:furtherreading}

The material given in this tutorial is based on~\cite{nonterm,interact}.
The former presents four trace-based coinductive operational semantics
for While, big-step and small-step relational semantics 
and big-step and small-step functional semantics, 
and prove their equivalence in the constructive setting of Coq.
The latter looked at a constructive account of interactive
resumptions, their important properties such as delay bisimilarity
and responsiveness, and gave several big-step operational semantics
for interactive While. 

Coinductive functional semantics similar to ones given in the tutorial
have appeared in the works of J. Rutten 
and V. Capretta~\cite{Rut:notcwb,Cap:genrct}.
J. Rutten gave in coalgebraic term 
a delayed state semantics for While, i.e.,
a semantics that, for a given statement-state pair, returns
a possibly infinitely delayed state. 
V. Capretta also looked at a delayed state based semantics
in his account of general recursion in constructive type theory.
Central to him was the realization that the delay type constructor
is a monad. 

A general categorical account of small-step semantics has been
given by I. Hasuo et al.~\cite{HJS:gentsc}.

Similar ideas also found applications in constructive formalization of
domain theory, in particular to compute least upper bounds. 
C. Paulin-Mohring~\cite{DomainKahn} developed a Coq
library for constructive pointed $\omega$-cpos and continuous
functions and gave semantics for Kahn networks based on them.
N. Benton et al.~\cite{DomainCoq} generalized her library to
treat predomains and a general lift monad, which are used 
to define denotational semantics for a simply-typed call-by-value
lambda calculus with recursion and an untyped
call-by-value lambda calculus. 

\section{Exercise}

1. Extend While with the statement \verb#repeat Stmt Bexp#,
and adapt the interpreter accordingly.\\[1ex]
2. Extend While with the statement \verb#Stmt :||| Stmt#,
and adapt the interpreter accordingly. \\
(\verb#stmt0 :||| stmt1# non-deterministically chooses to run either 
of \verb#stmt0# or \verb#stmt1#.)\\[1ex]
3. Write interactive programs that produce delay-bisimilar resumptions. \\[1ex]
4. Give resumptions that are delay-bisimilar, and prove that they are
indeed delay-bisimilar by finding a delay-bisimulation. 

\paragraph{Acknowledgments}
The author's research was supported by the
European Regional Development Fund (ERDF) through the Estonian Centre
of Excellence in Computer Science (EXCS).

\bibliographystyle{eptcs}
\bibliography{references}

\end{document}